\documentclass{aastex}          
\usepackage{spr-astr-addons}    
\usepackage{natbib}             
\usepackage{multirow}
\usepackage{bigstrut}

\newcommand{\rmn}{\textrm}
\begin{document}

\title{H$\alpha$ and UV luminosities and star formation rates in a large sample of luminous compact galaxies}

\shortauthors{S. L. Parnovsky et al.}

\shorttitle{H$\alpha$ and UV luminosities and SFRs in luminous compact galaxies}

\author{S. L. Parnovsky}
\affil{Astronomical Observatory of Taras Shevchenko Kyiv National University\\
Observatorna str., 3, 04058, Kyiv, Ukraine\\
tel: +380444860021, fax: +380444862191\\ e-mail:par@observ.univ.kiev.ua}
\email{par@observ.univ.kiev.ua}

\author{I. Y. Izotova}
\affil{Astronomical Observatory of Taras Shevchenko Kyiv National University\\
Observatorna str., 3, 04058, Kyiv, Ukraine\\
tel: +380444860021, fax: +380444862191\\ e-mail:izotova@observ.univ.kiev.ua}
\email{izotova@observ.univ.kiev.ua}

\author{Y. I. Izotov}
\affil{Main Astronomical Observatory\\
Zabolotnoho str., 27, 03680, Kyiv, Ukraine\\
tel: +380445264771, fax: +380445262147\\ e-mail:izotov@mao.kiev.ua}
\email{izotov@mao.kiev.ua}

\begin{abstract}
We present the results of a statistical study of the star formation rates (SFR)
derived from the {\sl Galaxy Evolution Explorer} ({\sl GALEX}) observations in
the ultraviolet continuum and in the H$\alpha$
emission line for a sample of about 800 luminous compact galaxies (LCGs).
Galaxies in this sample have a compact structure and
include one or several regions of active star formation. Global galaxy
characteristics (metallicity, luminosity, stellar mass) are intermediate
between ones of the nearby blue compact dwarf (BCD) galaxies and
Lyman-break galaxies (LBGs) at high
redshifts $z>$ 2 - 3. SFRs were corrected for interstellar extinction which
was derived from the optical Sloan Digital Sky Survey (SDSS) spectra.
We find that SFRs derived from the galaxy luminosities in the far
ultraviolet (FUV) and near ultraviolet (NUV) ranges vary in a wide range
from 0.18 $M_\odot$ yr$^{-1}$ to 113 $M_\odot$ yr$^{-1}$
with median values of 3.8 $M_\odot$ yr$^{-1}$ and 5.2 $M_\odot$ yr$^{-1}$,
respectively. Simple regression relations are found for luminosities
$L($H$\alpha )$ and $L$(UV) as functions of the mass of the young stellar
population, the starburst age, and the galaxy metallicity.
We consider the evolution of
$L$(H$\alpha$), $L$(FUV) and $L$(NUV) with a starburst age and introduce new
characteristics of star formation, namely the initial H$\alpha$, FUV and NUV
luminosities at zero starburst age.
\end{abstract}

\keywords{Galaxies: irregular --- Galaxies: luminosity function, mass function
--- Galaxies: starburst --- Galaxies: star formation --- Galaxies: statistics
}

\section{Introduction}\label{s:Introduction}

\citet{C09} first draw attention to
galaxies at redshifts $z$ = 0.112 - 0.360 which
were named ``green peas'' because of their compact structure
and green colour on the $gri$ composite Sloan Digital Sky
Survey (SDSS) images. Specific colours of these galaxies are mainly
caused by the very strong [O {\sc iii}] $\lambda$5007\AA\ optical
emission line. The equivalent widths EW($\lambda$5007)
of this line in ``green peas''
redshifted into the SDSS $r$ band can be as high as $\sim$ 1000\AA,
resulting in a green colour on SDSS
images. \citet{C09} studied a sample of 251 colour-selected
galaxies. Some of the galaxies from this sample are active galactic
nuclei (AGN). However, most of ``green pea'' galaxies are found to be
strongly star-forming ones with high star formation rates (SFR) of $\sim$
10 $M_\odot$ yr$^{-1}$. These galaxies are characterised by low
metallicity, stellar mass of $M_*$
$\sim$ 10$^{8.5}$ - 10$^{10}$ $M_\odot$, high specific SFR (SSFR)
(up to $\sim$ 10$^{-8}$ yr$^{-1}$) which place them
between nearby blue compact dwarf (BCD) galaxies and high-redshift
($z > 2 - 3$) UV-luminous Lyman-break galaxies \citep[LBGs, see
][for a review]{G02}. The available {\sl Hubble Space Telescope}
({\sl HST}) high-angular resolution images of a few ``green peas'' reveal
complex morphology on small spatial scales with several regions of
star formation and an extended stellar component likely consisting
of older stars \citep{C09,A12}. \citet{C09} suggested that ``green pea''
galaxies may be occurrences of the star formation mode prevailing
in the early Universe. This galaxy class therefore may provide an
excellent opportunity to understand in great detail many processes
under physical conditions approaching to those in high-redshift
galaxies.

The oxygen and nitrogen chemical abundances in star-forming ``green peas''
were studied by
\citet{A10}. These galaxies are revealed to be genuine metal-poor
galaxies with mean oxygen abundances of $\sim$ 20\% solar. The N/O ratios are
found to be unusually high for galaxies of the same metallicity. Detailed
study lead \citet{A10} to the conclusion that known general properties
of ``green peas'', namely high SSFR, extreme
compactness and stellar mass, seem to be uncommon in the nearby universe,
suggesting a short and extreme phase of their evolution. The possible action
of both recent and massive interaction-induced inflow of gas, as well as
selective metal-rich gas loss driven by supernova winds are discussed here as
main drivers of the starburst activity in ``green peas'' and their oxygen
and nitrogen abundances.

The first direct radio detection with low frequency Giant
Metrewave Radio Telescope (GRMT) observations and discussion of
the ``green peas'' properties comprising properties of a new class
of sub-mJy sources were reported by \citet{C12}. It was shown that
this detection may imply large magnetic fields ( $ \gtrsim 30 \,
\mu $G) in ``green peas'' under reasonable assumption about cosmic
ray diffusion and total energy consideration. \citet{C12}
concluded that seed fields were amplified
significantly (up to $\mu$G) because of turbulence as
protogalactic and similar structures formed.

    Detailed examination of a large sample of 803 star-forming luminous
compact galaxies (LCGs) in the redshift range $z$ = 0.02 - 0.63
was carried out by \citet{I11}. These galaxies were selected
from the SDSS Data Release 7 (DR7) \citep{A09}
and comprise a complete spectroscopic SDSS sample of strongly
star-forming LCGs with reliably derived chemical abundances.
Their global properties
are similar to those of the star-forming ``green pea''
galaxies. However, in contrast to ``green pea'' galaxies, the LCGs are
selected on the base of the both their spectroscopic and
photometric properties. Applied selection results in a $\sim$ 10
times larger sample, with galaxies spanning a redshift range about
$\sim$ 2 times larger as compared to ``green pea'' sample
\citep{C09}. For LCGs, the oxygen abundances 12 + log O/H are
found to be in the range 7.6 - 8.4 with the median value of $\sim$
8.11 confirming the results by \citet{A10} for a subset of the
``green pea'' sample of \citet{C09}. The ranges of oxygen
abundances and heavy element abundance ratios in LCGs are similar
to those of nearby low-metallicity BCDs. In the [O {\sc
iii}]$\lambda$5007/H$\beta$ vs. [N {\sc
ii}]$\lambda$6583/H$\alpha$ diagnostic diagram \citep{K03} the
LCGs are shown to occupy the region of high-excitation
star-forming galaxies. The SFRs, derived
from the H$\alpha$ line emission in the LCGs are revealed to vary
in the large range of 0.7 - 60 $M_\odot$ yr$^{-1}$, with a median
value of $\sim$ 4 $M_\odot$ yr$^{-1}$ which is about 3 times lower
as compared to star-forming LBGs at $z$ $\sim$ 3
\citep{P01}. The SSFR in LCGs is
extremely high and it varies in the range $\sim$ 10$^{-7}$ -
10$^{-9}$ yr$^{-1}$. All these properties imply that LCGs are
likely the closest local counterparts of the high-redshift LBGs and
Ly$\alpha$-emitting galaxies.

    \citet{G11} carried out the spectroscopic analysis of
HG 031203, the host galaxy of a long-duration gamma-ray burst
(GRB). The galaxy properties such as the oxygen
abundance 12 + log O/H  $=8.20\pm  0.03$, extinction-corrected
H$\alpha$ luminosity $L$(H$\alpha$) = 7.27$\times$10$^{41}$ erg
s$^{-1}$, stellar mass $M_*$ = 2.5$\times$10$^8$ $M_\odot$,
SFR(H$\alpha$) = 5.74 $M_\odot$ yr$^{-1}$ and
SSFR(H$\alpha$) = 2.3$\times$10$^{-8}$ yr$^{-1}$ in HG 031203 are found to be
in the range covered by the LCGs properties. This fact implies that
the LCGs with extreme star-formation, that also comprise ``green
peas'' as a subclass, may harbor GRB.

\citet{P12} analysed the oxygen and nitrogen
abundances derived from global emission-line SDSS spectra of
galaxies using the direct method based on the electron temperature
determination from the [O {\sc iii}] $\lambda$4363/($\lambda$4959 +
$\lambda$5007) emission-line flux ratio
and the two strong line O/N and N/S
calibrations. Three samples of objects were compared, including the
sample of ``green pea" galaxies by \citet{C09} with the detected
[O {\sc iii}] $\lambda$4363\AA\ auroral line. \citet{P12} concluded
that the high nitrogen-to-oxygen abundance
ratios derived in some ``green pea" galaxies may be due to the
fact that their SDSS spectra are the ones of composite nebulae made
up of several components with different physical properties.

The local analogues of the strong Halpha Emitters (HAEs) dominated
the $z\sim 4$ LBG population are identified by
\citet{SC12}. Using the SDSS spectra
authors show that at $z<0.4$ only 0.04\% of galaxies are
classified as HAEs with equivalent widths EW(H$\alpha$) of $>500$
\AA, comparable to that of $z\sim 4$ HAEs. Local HAEs have lower
stellar masses and lower UV luminosities than the $z\sim 4$
HAEs. On the other hand, their H$\alpha$-to-UV luminosity ratios and
SSFRs are consistent with those of $z\sim 4$ HAEs indicating that the
local analogues are the scaled-down versions of high-$z$
star-forming galaxies. Compared to the previously studied
local Lyman-break analogs (LBAs) of the $z\sim 2$ LBGs which were selected
using rest-frame UV fluxes \citep{H05}, the local HAEs show similar UV
luminosity surface densities, but lower metallicities and lower stellar masses.
This supports the idea that local HAEs are less evolved galaxies than the
traditional LBAs. Local HAEs show a strong He {\sc ii} $\lambda$4686 \AA\
emission line in the stacked spectrum, implying a population of hot young
($<10$ Myr) massive stars, similar to that seen in some Wolf-Rayet galaxies.
The local HAEs also have properties similar to those of ``green pea" galaxies.

In present paper, we extend the study of the properties of the
``green peas'' by further analysis of about 800 LCGs by
\citet{I11}, selected from the SDSS DR7.
The selection criteria of galaxy sample are briefly
described in Section \ref{s:Sample}. The correction of LCG fluxes
for extinction is discussed in Section \ref{s:Ext}. In Section
\ref{s:Dep} we carry out the statistical investigation of
dependence of galaxy luminosities on other LCG characteristics. In
Section \ref{s:SFR} we discuss star formation rates of LCGs. The
SFRs are derived from the extinction-corrected luminosities
$L$(H$\alpha$), $L$(FUV) and $L$(NUV) in the H$\alpha$ emission
line, the far ultraviolet (FUV) and the near ultrabviolet (NUV) ranges from
{\sl Galaxy Evolution Explorer} ({\sl GALEX}) observations.
The luminosity function for LCGs is discussed in Section \ref{s:LF}. We
summarise our results in Section \ref{s:Sum}.  We assume
$H_0$ = 75 km s$^{-1}$ Mpc$^{-1}$ for distance estimates.

\section{Sample selection, observational data}\label{s:Sample}

It is noted in Section \ref{s:Introduction}, that LCGs, in contrast
to ``green pea'' galaxies \citep{C09},
are selected on the base of the both their spectroscopic
and photometric properties. Selection criteria and LCGs sample
properties are described in detail by \citet{I11}.
Briefly, these criteria are as follows :

-- the extinction corrected luminosity of the H$\beta $
emission line is greater than $L$(H$\beta$) = 3$\times$10$^{40}$ erg s$^{-1}$;

-- the equivalent width of the H$\beta$ emission line is high,
EW(H$\beta$) $\geq $ 50\AA. This criterion leads to selection
only objects with strong emission lines in their spectra and thus the ones
containing young starbursts with ages 3~-~5 Myr;

-- only galaxies with well-detected [O {\sc iii}] $\lambda $4363 {\AA}
emission line in their spectra, with a flux error less than
50~{\%}, are selected. This criterion allows an
accurate abundance determination using the direct method;

-- only the star-forming galaxies were selected. Galaxies with obvious
evidence of Seyfert~2 features are excluded;

-- galaxies on their SDSS images are nearly compact at low
redshifts and unresolved at high redshifts. Their typical angular sizes
are less than 10\arcsec.

\citet{I11} used all LCG spectra and Monte Carlo simulations to fit
spectral energy distributions
in the wavelength range $\lambda$$\lambda$3800 -- 9200\AA. As for star
formation history they assumed a single young burst with the age which is
varied in the range $<$ 10 Myr, and a continuous star formation with a
constant star formation rate, which started at the lookback time $t_1$ and
finished at the lookback time $t_2$ $<$ $t_1$. Parameters $t_1$ and
$t_2$ are varied in the range 10 Myr -- 13 Gyr. The contribution of gaseous
continuum in LCGs is very high, therefore it was fitted first using
equivalent width EW(H$\beta$) of the H$\beta$ emission line and subtracted
from the observed spectrum prior fitting of the stellar continuum. The masses
of the young and old stellar populations, the age
of the young burst and parameters for the old stellar population $t_1$ and
$t_2$ were parameters of fitting. More details of fitting can be found in
\citet{I11}. We use the results of modelling obtained by \citet{I11}.

We use the {\sl GALEX} Medium Imaging Survey (MIS) and
All-sky Imaging Survey (AIS) data (see http: //galex.stsci.edu/GR4) to
estimate the galaxy UV SFR for the LCGs sample. {\sl GALEX} is a NASA
Small Explorer mission performed the all sky ultraviolet survey in two
bands: far-UV (FUV, $\lambda_{\rm eff}$=~1528\AA), and near-UV
(NUV, $\lambda_{\rm eff}$ =~2271\AA) \citep{M05}. MIS and AIS data contain
information on fluxes of $\sim $10$^{7}$ galaxies.
The prime goal of {\sl GALEX} is to study star formation in galaxies and its
evolution with time. The major science objectives and characteristics of
{\sl GALEX}, and of surveys are described by \citet{Ma05} and
\citet{M05}.

We matched the {\sl GALEX} data and the sample of LCGs \citep{I11} and
extracted FUV
and NUV fluxes from the {\sl GALEX} MIS and AIS database. These data combined
with the NASA/IPAC Extragalactic Database (NED) data provide the determination
of the galaxy UV luminosities.

We excluded the galaxies with the UV flux errors exceeding 50\% and the data
for the multiple UV sources within the aperture of $\sim$30\arcsec.

\section{Correction for extinction}\label{s:Ext}

Because the radiation of galaxies is reduced by dust extinction, we applied
reddening corrections to H$\alpha$ and UV band fluxes using \citet{C89}
reddening
law. Adopting the $R(V)$-dependent extinction law from \citet{C89}
with $R(V)$ = $A(V)/E(B-V)$ = 3.1, we obtain
$A$(H$\alpha$) = 2.54$\times$$E(B-V)$ in H$\alpha$,
$A$(FUV) = 8.15$\times$$E(B-V)$
in the FUV band and $A$(NUV) = 9.17$\times$$E(B-V)$ in the NUV
band. The extinction coefficient $C$(H$\beta$),
reddening $E(B-V)_{\rm SDSS}$, and the equivalent width of underlying
stellar hydrogen absorption lines were obtained by \citet{I11} from the
hydrogen Balmer decrement in the redshift-corrected spectra.
All hydrogen line fluxes were corrected for both the reddening and underlying
stellar absorption. For comparison,
we also use the reddening $E(B-V)_{\rm NED}$ from
the NED which was obtained from the Milky Way reddening maps by \citet{S98}.
The $E(B-V)_{\rm SDSS}$ and $E(B-V)_{\rm NED}$ differ because the
former quantity is the total reddening along the line of sight which includes
extinction from both the Milky Way and the studied galaxy, while the latter
quantity is the reddening in the Milky Way only.

\citet{I11} derived $E(B-V)_{\rm SDSS}$ = 0 for 65 out of 803 galaxies. For
some other galaxies they obtained $E(B-V)_{\rm SDSS}$ $<$ $E(B-V)_{\rm NED}$.
To correct galaxy fluxes for extinction we use the $E(B-V)_{\rm SDSS}$ for all
LCGs.
Alternatively, we analyse all the UV data adopting $E(B-V)_{\rm SDSS}$ if
$E(B-V)_{\rm SDSS}$ $>$ $E(B-V)_{\rm NED}$ and $E(B-V)_{\rm NED}$ if
$E(B-V)_{\rm SDSS}$ $<$ $E(B-V)_{\rm NED}$.
The difference between these two approaches
is small because of the low extinction in LCGs and does not influence
appreciably our results and conclusions (see Sect. \ref{s:Dep}).

We derive mean reddenings $E(B-V)_{\rm SDSS}$ of 0.133 and 0.134 for LCGs
detected in the FUV and NUV ranges, respectively. We also find that the mean
reddening difference $E(B-V)_{\rm SDSS}$ $-$ $E(B-V)_{\rm NED}$ is 0.106.
This difference is the rough
average internal galaxy reddening which can be used for galaxies without
spectroscopic estimates of reddening.
In principle, the correction for the Milky Way and intrinsic
reddening should be done separately. This is because the Milky Way correction
has to be applied to the fluxes at the observed wavelengths, while the
correction for the intrinsic reddening should be applied to the fluxes at
the redshift-corrected wavelengths. However, ignoring of the separate
correction for the Milky Way and intrinsic reddening would introduce very
small additional uncertainties in the extinction-corrected fluxes. All LCGs
are located at high galactic latitudes where the Milky Way extinction
is very low, with the mean $E(B-V)_{\rm NED}$ of $\sim$ 0.03. Most of LCGs are
also low-redshift galaxies with $z$ $<$ 0.2. Only few galaxies are at
redshifts $z$ $>$ 0.3. Therefore, the
difference between the extinction correction of the flux applied with the
observed wavelength and the redshift-corrected wavelength for the galaxy with
the redshift $z$ = 0.3 and adopting $E(B-V)_{\rm NED}$=0.03 is $\la$ 10\% in
the FUV and NUV bands, and $\la$3\% for the H$\beta$ and H$\alpha$ emission
lines. This difference for galaxies with lower redshifts is lower, e.g. it is
only $\sim$ 5\% in the FUV and NUV bands and $\sim$ 2\% for H$\beta$ and
H$\alpha$ for the galaxy with $z$ = 0.2. Furthermore, Milky Way extinction
maps by \citet{S98} are obtained for large apertures of 6\arcmin, which are
much larger than the angular sizes of LCGs. Therefore, the extinction
derived from the \citet{S98} maps may not correspond
to the real extinction in the direction on the galaxy if small-scale spatial
extinction variations are present. These are reasons why we do not separate
correction for the Milky Way and intrinsic reddening and use in a subsequent
analysis $E(B-V)_{\rm SDSS}$ derived by \citet{I11} from the
hydrogen Balmer decrement.

Accordingly to \citet{I11} we split our sample into two subsamples of 276 ``regular'' galaxies
with the round shape and 519 ``irregular'' galaxies with some sign of disturbed
morphology suggesting the presence of two or more star-forming regions and
their interaction. For these two subsamples we obtained
$E(B-V)_{\rm SDSS}$ $-$ $E(B-V)_{\rm NED}$ of 0.081 and 0.120, respectively.
Using the Student criterion we derived $t$ = 4.98. This value suggests that the
probability of the statistically significant difference of
$E(B-V)_{\rm SDSS}$ $-$ $E(B-V)_{\rm NED}$ for two subsamples is greater than
99.9\%. Therefore, extinction in galaxies with the non-round morphology
is higher.
On the other hand, we do not find
tight correlation between $E(B-V)_{\rm SDSS}$ $-$ $E(B-V)_{\rm NED}$ and
heavy element abundances. Apparently, the extinction is determined not
only by the dust mass which is expected to be higher in galaxies with higher
metallicity, but also by the spatial distribution of dust. This distribution
seems to be different in ``regular'' and ``irregular'' galaxies.

We correct the galaxy fluxes for extinction according to
$I(\lambda)=F(\lambda)\times 2.512^{A(\lambda)}$, where $F(\lambda)$
and $I(\lambda)$ are the observed and the corrected fluxes, respectively.
The extinction-corrected {\sl GALEX} FUV and NUV fluxes from LCGs are
nearly three times higher than the observed fluxes. In addition, H$\alpha$
fluxes were corrected for an aperture comparing the total galaxy apparent
magnitude $m$ and the magnitude $m$(3\arcsec) inside the SDSS spectroscopic
aperture of 3\arcsec\ in a certain SDSS band depending on the galaxy redshift.
We compare SDSS magnitudes $m$ = $r$ and $m$(3\arcsec) = $r$(3\arcsec) for
galaxies with redshifts $<$ 0.04, $i$ and $i$(3\arcsec) for galaxies
with redshifts in the range 0.04 -- 0.26, and $z$ and $z$(3\arcsec) for
galaxies with redshifts $\geq$ 0.26. Then, the aperture H$\alpha$
flux correction is $A$=2.512$^{m-m(3\arcsec)}$, where $m$ = $r,i,z$
depending on the galaxy redshift.

\section{Relations between galaxy luminosities and other global
characteristics}\label{s:Dep}

For each galaxy we calculated its H$\alpha$, FUV and NUV luminosities.
We use the regression analysis to study
a dependence of the LCG luminosities on other their characteristics. To
provide a simple way for
comparing $L($H$\alpha)$, $L$(FUV) and $L$(NUV)
we use some other parameters being proportional to them.
Namely, we use the calibration for SFRs averaged over the reasonable
timescale for different SFR tracers and defined by \citet{K98} as
\begin{equation}\label{eqn:1}
\rmn{SFR}=k \times L.
\end{equation}
The conversion factors $k$ between the SFR and the
$L$(H$\alpha$), $L$(FUV) and $L$(NUV) in Eq. \ref{eqn:1} are derived using the
evolutionary synthesis models. The coefficient $k$ depends on the time scale of
star formation, initial mass function (IMF) and galaxy metallicity.
Adopting the solar metallicity, the IMF with the power-law index 2.35 and mass
limits of 0.1 and 100 $M_\odot$ \citep{S55}, SFR in $M_{\odot}$ yr$^{-1}$,
$L($H$\alpha)$ in erg s$^{-1}$, $L$(FUV) and $L$(NUV) in erg s$^{-1}$ Hz$^{-1}$
\citet{K98} obtained the coefficient $k$ of $7.9 \times 10^{-42}$ for
the H$\alpha$ luminosity and
$1.4 \times 10^{-28}$ for the FUV and NUV luminosities.

There are some other modifications of Eq. \ref{eqn:1}. In particular,
\citet{K2009} proposed a composite SFR calibration based on the luminosities of
both the H$\alpha$ emission line and the UV continuum, adopting the IMF by
\citet{Kr} and obtained $k$ = $5.5 \times 10^{-42}$. For clarity, we will use
the values of $k$ derived by \citet{K98} for the solar metallicity and Salpeter
IMF.
Detailed review and analysis of SFR calibrations based on the
H$\alpha$ $\lambda$6563\AA\ and
[O {\sc ii}] $\lambda$3727\AA\ emission lines, far infrared and ultraviolet
continua are given by \citet{K98} and \citet{Ca12}. 

While the calibration defined by Eq. \ref{eqn:1} holds for continuous or
quasi-continuous star formation which is common in the big galaxies with
frequent starbursts, the situation is more
complicated in the dwarf galaxies like LCGs with the strong and rare bursts of
star formation. In these systems with the instantaneous bursts the
time interval of the star formation is not well defined, and the observed
H$\alpha$ luminosity strongly decreases on a time
scale of a few Myr.
Similar conclusions in lesser extent can be drawn for the SFRs derived
from the FUV-band and the NUV-band luminosities. However,
again for the clarity, we adopt that Eq. \ref{eqn:1} can be applied for
the determination of SFR in LCGs.

A set of galaxy parameters includes the primary parameters
obtained directly from its SDSS spectrum, such as the redshift, the H$\alpha$
flux and the chemical element abundances. For each galaxy we also use
secondary parameters obtained by modelling galaxy spectral energy
distribution, following \citet{I11}, namely the masses of
young $M$(young) and old $M$(old) stellar populations, the total mass of a
stellar population $M_{*}$, the age of a starburst $t$(young), and the lower
($t_2$) and upper ($t_1$) limits for the age of old stars. All these
parameters are distance-independent.
Nevertheless, some nearby galaxies from our sample have angular
diameters larger than the 3\arcsec\ SDSS aperture. Neglecting an aperture
correction for these galaxies would result in the underestimation of stellar
population masses.
To estimate the proper masses we used the aperture correction similar to the
one used for luminosities. The luminosities of galaxies with angular diameters
greater than 3\arcsec\ without this correction deviate substantionally from
regression relations used for approximation of their SFRs as a function of
their luminosities. These deviations vanish after taking into account aperture
corrections.

We consider all above-mentioned parameters. However, as
one can see below, only two secondary parameters,
namely $M$(young) and $t$(young) have the statistically significant impact
on the luminosity. Hereafter we will use the brief notations
$m$ $\equiv$ $M$(young) and $T$ $\equiv$ $t$(young) for the mass of the
young stellar population and its age, respectively.

\begin{figure}[tb]
\includegraphics[width=\columnwidth]{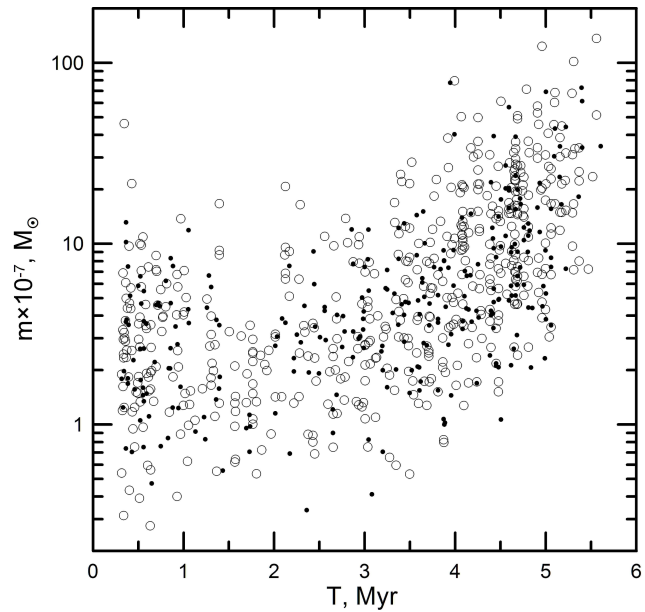}
\caption{Mass of the young stellar population $m$ vs. the starburst age
$T$. Dots and open circles correspond to subsamples of ``regular'' and
``irregular'' galaxies, respectively}
\label{fig:1}
\end{figure}

Consider distributions of some secondary parameters. The values of the
starburst age $T$ are distributed rather uniformly over the interval
$T<5$ Myr. Galaxies with $T \ga 5.6$ Myr are not included in our sample due to
the criterion EW(H$\beta$) $>$ 50\AA. This selection is also
resulted in a relative decrease of the number of galaxies with $5<T<5.6$ Myr.
The distribution of $m$ depends on $T$. This is illustrated in Figure
\ref{fig:1}. One can see that the mean value of $m$ for $T>3.2$ Myr increases
with the increase of $T$ with the best fit
$\log(m/10^7 M_\odot)=-0.57 \pm 0.22 +
(0.33 \pm 0.05)\times T$(Myr) for the subsample of ``regular'' galaxies
and $\log(m/10^7 M_\odot)=-0.75 \pm 0.17
+(0.40 \pm 0.04) \times T$(Myr) for the subsample of ``irregular'' galaxies.

Note that both values $m$ and $T$ were calculated assuming a single
star-forming region in the galaxy, while several regions of star formation
with different $m$ and $T$ sometimes are observed in the galaxies.
In this case we cannot rely on single values of
$m$ and especially $T$, the latter value would tend to be larger.
To prove that we consider a case of the two star-forming regions with equal
stellar masses, one is very
young and another is older. The young star-forming region would
dominate in the H$\alpha$ luminosity because of the strong dependence
of the flux of ionising radiation on a starburst age. On the other
hand, the intensity of the optical continuum is less sensitive to
the starburst age, therefore both star-forming regions equally contribute
to the optical continuum, resulting in lower EW(H$\alpha$) as compared
to the case with a single young burst. Consequently, this
would result in a larger $T$ because it is determined mainly by
EW(H$\alpha$). It is natural to expect that galaxies with larger masses of
young stellar population with higher probability consist of several
regions of star formation at different evolutionary stages. Therefore they
would tend to have larger
$T$ as compared to the galaxies with smaller masses of the young stellar
population. That is why these galaxies concentrate in the upper right corner in
Figure \ref{fig:1}.

We will show later that the ratio $L/m$ decreases exponentially with
increasing $T$ if $T>3.2$ Myr. As a result, the lower right corner in
Figure \ref{fig:1} is
empty because the sample is flux-limited
$L$(H$\beta$) $>$ 3$\times$10$^{40}$ erg s$^{-1}$. Galaxies with low $m$ and
$T>4$ Myr have luminosities below the threshold and do not enter the sample.

Our goal is to find simple but statistically significant dependences of the
galaxy luminosity on other primary and/or secondary galaxy parameters. First,
we search for a set of parameters to which the galaxy luminosity is the most
sensitive. Later, we will find the best formulae to describe these dependences
and analyse them.

At first we do not take into account galaxy metallicities and analyse
linear dependences of their luminosities on other parameters and their
combinations. Regressors were chosen accordingly to the statistical Fisher test
\citep{ref:F}. We reject the regressors with statistical significance
below the threshold of 99.95\% and consider the regressions which are
good for all six subsamples by selecting 3 wavelengths
(H$\alpha$, FUV and NUV) and 2
morphologies - ``regular'' and ``irregular''. These regressions have the form
\begin{equation}\label{eqn:2}
\mathrm{SFR}=C_1+C_2m+C_3mT^2+C_4m^2.
\end{equation}
The values, the errors and the statistical
significances of the coefficients $C_i$ obtained by the least
squares method (LSM) are shown in Table \ref{tbl:1} for $m$ expressed in solar
masses and $T$ in yr. The root mean square (RMS) standard deviation  $\sigma$
and the number $N$ of the galaxies in each subsample are also shown in the
Table.
Significances are characterised by the value $F$
obtained by Fisher's test. The critical $F$ values, corresponding
to the statistical significances of 90, 95, 97.5, 99, 99.5, 99.9
and 99.95\% are equal to 2.71, 3.84, 5.02, 6.64, 7.88, 10.83 and
12.10, respectively.
One can see from Table \ref{tbl:1} that the
significances of regressors $C_2$ and $C_3$ in Eq. \ref{eqn:2} are
higher than 99.95\%. For the regressor $C_4$ we choose the
threshold value $F>10$. Only for one subsample the threshold attains a
higher value. For all other subsamples we assume $C_4=0$ and
indicate in parentheses the value of $F$ for the case $C_4\ne 0$.

\begin{table*}[tb]
\caption{Coefficients $C_i$ in Eq. \ref{eqn:2} and their significance values $F$ for
different galaxy subsamples}
\begin{tabular}{lcccccc}
\hline
Subsample&$N$&$\sigma$&$C_1$($F$)&$C_2\times 10^8$($F$)&$C_3\times 10^{22}$($F$)&$C_4\times 10^{18}$($F$)\\
\hline
\multicolumn{7}{c}{a) the case with $C_1$ $\ne$ 0} \\ \hline
1. H$\alpha$,``regular''    &$276$&$3.4$&$1.15\pm 0.35(10.9)$ &$18.3\pm 0.8( 528)$&$-43.1\pm 2.6( 268)$&$-33.8\pm 8.0(17.8)$\\
2. H$\alpha$,``irregular''  &$519$&$4.0$&$1.41\pm 0.24(35.3)$ &$17.0\pm 0.4(1602)$&$-45.5\pm 1.6( 766)$&$-(0.7)$            \\
3. FUV,``regular''          &$213$&$4.8$&$0.19\pm 0.45( 0.2)$ &$12.4\pm 1.0( 149)$&$-22.1\pm 4.0(  30)$&$-(3.9)$            \\
4. FUV,``irregular''        &$418$&$4.4$&$1.29\pm 0.31(17.5)$ &$ 8.0\pm 0.6( 159)$&$-14.2\pm 2.4(  34)$&$-(0.4)$            \\
5. NUV,``regular''          &$233$&$5.6$&$0.42\pm 0.50( 0.7)$ &$14.5\pm 1.1( 167)$&$-20.7\pm 4.4(  22)$&$-(2.3)$            \\
6. NUV,``irregular''        &$435$&$7.8$&$0.27\pm 0.53( 0.3)$ &$14.8\pm 1.1( 180)$&$-23.5\pm 4.2(  31)$&$-(8.7)$            \\ \hline
\multicolumn{7}{c}{b) the case with $C_1$ $=$ 0} \\ \hline
1. H$\alpha$,``regular''    &$276$&$3.5$&$0$                  &$20.2\pm 0.6(1275)$&$-46.3\pm 2.5( 343)$&$-49.2\pm 6.6(55.2)$\\
2. H$\alpha$,``irregular''  &$519$&$4.1$&$0$                  &$18.4\pm 0.4(2592)$&$-49.8\pm 1.5(1062)$&$-(2.0)$            \\
3. FUV,``regular''          &$213$&$4.8$&$0$                  &$12.6\pm 0.8( 227)$&$-22.8\pm 3.6(  40)$&$-(1.8)$            \\
4. FUV,``irregular''        &$418$&$4.5$&$0$                  &$ 9.6\pm 0.5( 352)$&$-19.3\pm 2.1(  82)$&$-(0.4)$            \\
5. NUV,``regular''          &$233$&$5.6$&$0$                  &$15.1\pm 0.9( 264)$&$-22.3\pm 4.0(  32)$&$-(0.5)$            \\
6. NUV,``irregular''        &$435$&$7.8$&$0$                  &$15.2\pm 0.9( 303)$&$-24.6\pm 3.6(  47)$&$-(7.1)$            \\
\hline

\end{tabular}\label{tbl:1}
\end{table*}

Consider an implication of Eq. \ref{eqn:2}. One
would expect that SFRs vanish at the low-mass limit $m=0$.
However, Eq. \ref{eqn:2} (case (a) in Table \ref{tbl:1}) implies
that SFR is equal to non-zero $C_1$ at this limit with a large
statistical significance for the three subsamples out of six. We
assume that this is due to the uncertainties in the estimation of $m$.
We suggest that this statistical effect resembles the well-known Malmquist
bias. A similar effect was studied in connection with the
large-scale collective galaxy motion \citep{ref:ParPar08}. To
verify this hypothesis we performed some Monte Carlo simulations.
For these simulations we need many generated mock catalogues, preferably with
the distribution of the parameters similar to that in real subsamples.

First, we adopt the values of $m$ and $T$ from the corresponding real
subsamples. Then, we set $C_1 = 0$ and calculate the coefficients $C_2$, $C_3$,
and $C_4$ (if the last one is statistically significant) for the regression
relation Eq. \ref{eqn:2} using the LSM. These coefficients are
shown in Table \ref{tbl:1}, case (b).
After that, we derive the SFR values for the case (b) from
Eq. \ref{eqn:2} with $C_1 = 0$.
As a result we obtain the initial ``unbiased'' set of $m$,
$T$ and SFR values. Note that the dependence of SFR on $m$ and $T$ in this
sample is functional, not statistical. The real SFR data are taken into
account only indirectly via the set of coefficients $C_2$, $C_3$ and $C_4$.
Naturally, if we apply the LSM with the regression defined by Eq. \ref{eqn:2}
to this data we obtain coefficients in the corresponding row of Table
\ref{tbl:1}, case (b), but with $\sigma=0$.

Next, using the Monte Carlo technique we add a noise, i.e. random errors to
the unbiased values of SFR or $m$. We find that random normal errors in SFR
values result
in the nonshifted distributions of $C_i$ values obtained by the
LSM. In this case a random value for $C_1$ has the low
statistical significance. A completely different situation arises
when random errors in $m$ are considered. In this case the
distributions of $C_i$ values are shifted relative to the
``unbiased" ones and we obtain a non-zero value for $C_1$, sometimes with
the large false statistical significance. Other coefficients tend to attain
values nearer to zero if the noise increases.

\begin{table*}[tb]
\caption{Coefficients $C_i$ in Eq. \ref{eqn:2} for the subsample No. 1
(Table \ref{tbl:1}) obtained from
Monte Carlo simulations adopting ``unbiased'' values of SFR and $T$
and different amplitudes $s$ of the noise for $m$ according to Eq. \ref{eqn:3}}
\centering{
\begin{tabular}{ccccccc}
\hline
$s$&$C_1$&$\bigstrut C_2\times 10^8$&$C_3\times 10^{22}$&$C_4\times 10^{18}$&$\sigma$\\
\hline
$0.20$&$0.44$&$19.1$&$-43.1$&$-45.3$&$1.83$\\
$0.30$&$0.98$&$17.5$&$-39.0$&$-40.9$&$2.72$\\
$0.32$&$1.11$&$17.2$&$-38.1$&$-39.8$&$2.87$\\
$0.34$&$1.24$&$16.9$&$-37.2$&$-38.7$&$3.03$\\
$0.35$&$1.30$&$16.6$&$-36.5$&$-38.2$&$3.10$\\
$0.40$&$1.61$&$15.8$&$-34.3$&$-35.4$&$3.43$\\
\hline
\end{tabular}\label{tbl:2}
}
\end{table*}

Could the effect of uncertainties in the $m$ determination explain a
non-zero value of $C_1$ obtained from the real data? To prove this
suggestion we compare the values of $C_i$ in Table \ref{tbl:1}
(case (a)) with the ones obtained using Monte Carlo simulations.
For that, we use the ``unbiased'' SFR and $T$ values from the sample
calculated with the coefficients from case (b) in Table \ref{tbl:1}
and add a random noise to $m$. We obtain the ``biased'' value $m_{bias}$.
The distribution of $m_{bias}$ is expected to be
log-normal, therefore we added the noise using the equation
\begin{equation}\label{eqn:3}
m_{bias}=m(1+s \times \xi),
\end{equation}
where $\xi$ is a normally distributed random value with zero mean and unity
dispersion and $s$ characterises an amplitude of the noise.
Then we calculate $C_i$ values applying LSM to the set of ``unbiased'' SFR and
$T$ values and the values of $m_{bias}$. After repeating
this procedure 1000 times we obtain the mean values
and the distribution of $C_i$ as well as the mean value of $\sigma$.
The results for different $s$ are shown in Table
\ref{tbl:2} for the H$\alpha$ subsample No. 1 in Table \ref{tbl:1}.
One can see that the calculated values of $C_1$ in Table \ref{tbl:1} (case (a))
can be explained by the Monte Carlo model with $s=0.34$ (Table \ref{tbl:2}).
In addition to the shift of $C_1$ the noise in $m$ also affects the values of
other coefficients $C_i$. They are shifted closer to the
values from Table \ref{tbl:1} (case (a)). The value of modelled $\sigma$ is
smaller than that obtained from real data due to the contribution of SFR errors
in real data, in addition to errors in $m$.

We used
similar Monte Carlo simulations for other two subsamples with statistically
significant non-zero $C_1$ values (subsamples No. 2 and 4 in Table \ref{tbl:1})
and obtained similar results attained at $s=0.32$ and $s=0.38$, respectively.

Therefore, we adopt that the true preliminary regression for SFR
is Eq. \ref{eqn:2} with $C_1=0$ and non-zero values of $C_2$, $C_3$ and
sometimes $C_4$ (see case (b) in Table \ref{tbl:1}). The uncertainties in the
estimation of $m$ lead to the appearance of
the first regressor in Eq. \ref{eqn:2} with a false statistical significance.
Hereafter we will use only the models with $C_1=0$. Then we can rewrite Eq.
\ref{eqn:2} in the form
\begin{equation}\label{eqn:4}
\mathrm{SFR}/m=C_2+C_3T^2+C_4m,
\end{equation}
introducing a new value SFR/$m$ which
gives us a possibility to consider the dependence on $T$
regardless of the dependence on $m$.
The first term in Eq. \ref{eqn:4} is the main
one and its meaning is that SFR of a galaxy is approximately
proportional to the mass of the young stellar population. This is quite obvious
because radiation in the UV continuum and H$\alpha$ emission line is emitted mainly by
young O-stars.
We will return later to the consideration of the possible nonlinear dependence
of SFR on $m$.

\begin{figure}[tb]
\includegraphics[width=\columnwidth]{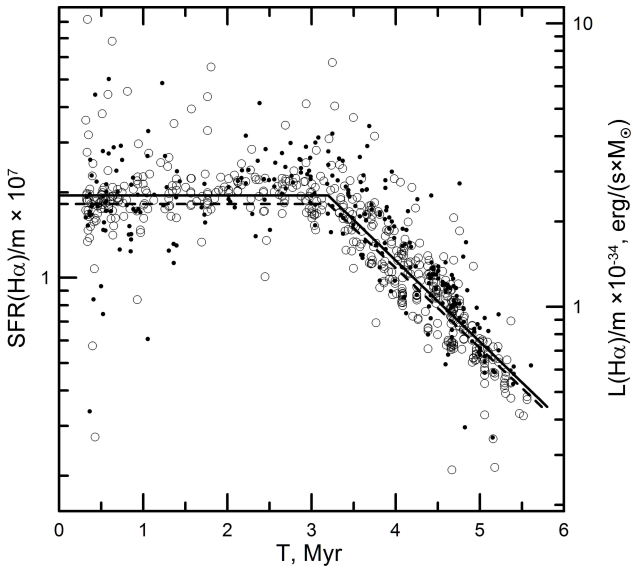}
\caption{Ratio $L$(H$\alpha$)/$m$ of the galaxy luminosity in the H$\alpha$ emission line to the mass $m$
of the young
stellar population vs. the age of the starburst $T$. Dots and open circles correspond
to subsamples of galaxies with ``regular'' and ``irregular'' shape. The solid and dashed
lines correspond to the best fits for the subsamples 1 and 2 with the mean galaxy metallicity
using the regression Eq. \ref{eqn:9}}
\label{fig:2}
\end{figure}

In Figure \ref{fig:2} we show the dependence of $L$(H$\alpha$)/$m$
$\propto$ SFR/$m$ on the starburst age $T$. It is seen
from the Figure that the ratio $L$(H$\alpha$)/$m$ is practically constant for
$T < T_0$=3.2 Myr and decreases practically exponentially for larger $T$.
We note that SFR(H$\alpha$), $m$, and $T$ are not directly
correlated because they are based on the different features
in the spectra: SFR(H$\alpha$) is not modelled and is derived from
the H$\alpha$ flux, $m$ and $T$ are modelled but they depend on
different features. The mass of young stellar population $m$ is determined
by the luminosity and the shape of the stellar continuum in the visible range,
while $T$ depends mainly on the equivalent width EW(H$\beta$) of the H$\beta$
emission line. In particular,
the equivalent width EW(H$\beta$) is lower by a factor of many times
for the burst with the age of 6 Myr as compared to that for the burst with the
age of 3 Myr, while the flux of the stellar continuum near H$\beta$ is
decreased by only $\sim$ 5\% \citep{L99}. Therefore, SFR(H$\alpha$), $m$, and
$T$ can be considered as independent parameters. The most plausible explanation
of the $L$(H$\alpha$)/$m$ -- $T$ relation is that
most massive stars with masses of $\sim$ 100 $M_\odot$,
producing most of the ionising radiation, disappear after
the starburst age $T$ $\sim$ 3.2 Myr, corresponding to their lifetimes.
It also implies that stars in LCGs are formed during very short time periods,
otherwise, in the case of extended bursts, $T_0$ would be greater than
$\sim$ 3.2 Myr. The best fit for $T > 3.2$ Myr is
$\log({\rm SFR}/m) = -5.61\pm 0.07+(-0.316\pm 0.016)\times T({\rm Myr})$ for the ``regular''
subsample and $\log({\rm SFR}/m) = -5.62\pm 0.05+(-0.326\pm 0.011)\times T({\rm Myr})$
for the ``irregular'' subsample.

Is this effect statistically significant? The values of the Fisher coefficients for linear
terms in regression relations for both subsamples exceed 398. For additional proof we performed
the Student test to compare the mean values of the  SFR/$m$ for $T<3.5$ Myr and $T>4$ Myr.
In all cases the Student test gives $t$ above 3.31, corresponding to the
statistical significance of 99.95\%.
The results are shown in Table \ref{tbl:3}, where $\sigma$ is the standard deviation. One can see that mean values of
SFR/$m$ for $T<3.5$ Myr are essentially larger than
those for $T>4$ Myr. This effect is more pronounced for
the H$\alpha$ radiation as compared to the FUV and NUV radiation.
This is because 1) more massive short-lived stars contribute to the ionising
radiation and 2) the luminosity of ionising radiation is much stronger
increased with the mass of a star as compared to the UV-radiation.
We note, however, that the ratio SFR/$m$ decreases more slowly in comparison
with the \citet{SV98} population synthesis models for young stellar
populations given the appropriate heavy element abundance.

\begin{table*}[tb]
\caption{Comparison of the average values of SFR/$m$ for different starburst
ages $T$ according to the Student's $t$-test}
\centering{
\begin{tabular}{lcccccccc}
\hline
\multirow{2}{*}{Sample}&\multicolumn{3}{c}{$T<3.5$ Myr}&&\multicolumn{3}{c}{$T>4$ Myr}&\multirow{2}{*}{$t$}\\ \cline{2-4} \cline{6-8}
&$N$&SFR/$m\times 10^8$&$\sigma\times 10^8$&&$N$&SFR/$m\times 10^8$&$\sigma\times10^8$&\\
\hline
1. H$\alpha$,``regular''    &$137$&$20.9$&$ 6.9$&&$100$&$8.8$&$3.3$&$17.8$\\
2. H$\alpha$,``irregular''  &$249$&$20.9$&$ 8.1$&&$215$&$7.9$&$3.0$&$23.5$\\
3. FUV,``regular''          &$104$&$12.4$&$ 8.7$&&$ 79$&$7.5$&$5.9$&$ 4.6$\\
4. FUV,``irregular''        &$198$&$13.8$&$11.8$&&$177$&$5.4$&$2.5$&$ 9.7$\\
5. NUV,``regular''          &$114$&$15.6$&$11.1$&&$ 88$&$9.7$&$7.5$&$ 4.5$\\
6. NUV,``irregular''        &$207$&$19.5$&$18.6$&&$181$&$8.4$&$4.9$&$ 8.2$\\
\hline
\end{tabular}\label{tbl:3}
}
\end{table*}

In accordance with the above discussion we introduce the function $f(T)$
\begin{equation}\label{eqn:5}
f(T)=\left\{
\begin{array}{rl}
1&\text{if $T<T_0$;}\\
\exp(-p\times(T-T_0))&\text{if $T>T_0$,}\\
\end{array}\right.
\end{equation}
where $T_0=3.2$ Myr and $p= 0.75$ Myr$^{-1}$. These values are the
preliminary ones obtained from Figure \ref{fig:2}. They will be improved later.
Using $f(T)$ and transforming Eq. \ref{eqn:4} we obtain the regression relation
\begin{equation}\label{eqn:6}
SFR= C_5mf(T)+C_4m^2
\end{equation}
and apply LSM to calculate the values and the errors of the coefficients.

The improved dependence of SFR/$m$ on $T$ (dashed and solid lines in
Fig. \ref{fig:2} correspond to best fits with the optimal values of $p$)
and the existence of the correlation between  $T$ and $m$
(Fig. \ref{fig:1}) lead us to the following results. For the H$\alpha$
``regular'' subsample the second term
becomes statistically insignificant and for H$\alpha$ ``irregular'' subsample
the second term becomes positive and statistically significant. The values
of RMS standard deviations become smaller than those from Eq. \ref{eqn:4}
because of the better approximation of SFR/$m$ on $T$.

\begin{figure*}[tb]
\includegraphics[width=\columnwidth]{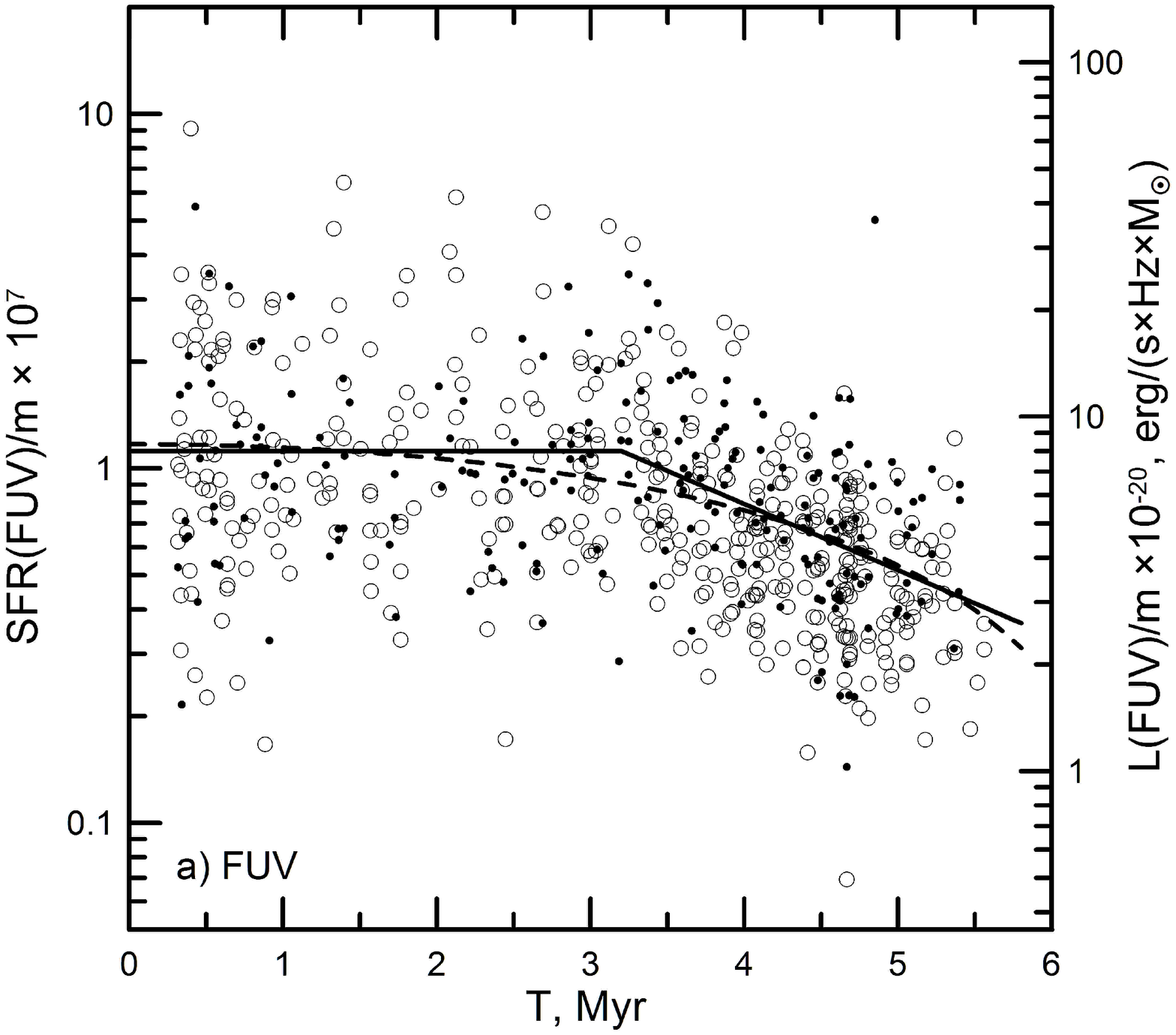}
\includegraphics[width=\columnwidth]{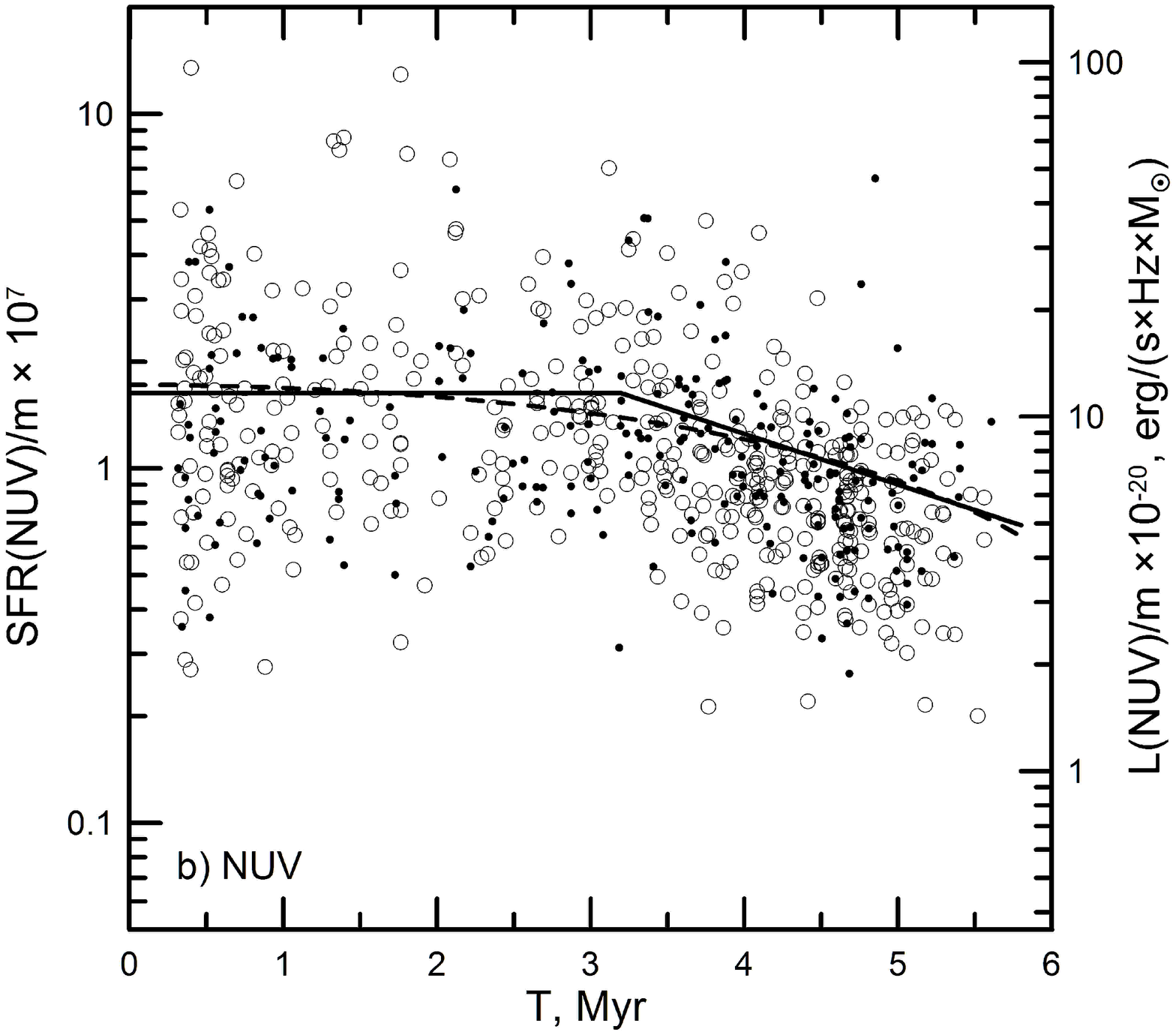}
\caption{Ratio $L$(UV)/$m$ of the luminosity in the UV range to the mass $m$
of the young
stellar population vs. the age of the starburst $T$ for the a) FUV and b) NUV radiation.
Dots and open circles correspond to subsamples of galaxies with ``regular'' and ``irregular''
shape. The solid and dashed lines show the best fits (Eq. \ref{eqn:9})
with $f(T)$ in Eq. \ref{eqn:5} and Eq. \ref{eqn:10}, respectively, and
adopting $[{\rm O}]=\left<[{\rm O}]\right>$}
\label{fig:3}
\end{figure*}

The dependences of $L$(FUV)/$m$ and $L$(NUV)/$m$ on $T$ are
more gently sloping and have a larger scattering of $L$(FUV)$/m$ and
$L$(NUV)$/m$ as compared to $L$(H$\alpha$)$/m$. The dependences $L$(FUV)$/m$
on $T$ and $L$(NUV)$/m$ on $T$
are shown in Figure \ref{fig:3}. The shape and application of
$f(T)$ to describe these dependences will be discussed below.
First of all we will use the function $f(T)$ for the H$\alpha$ radiation.

We now consider the dependence of $L$ on the galaxy metallicity. As a measure
of metallicity we choose the oxygen abundance [O] $\equiv$ 12 + log(O/H).
[O] was accurately derived by \citet{I11} for every galaxy in the LCG sample.
We adopt these values for the regression determination. To analyse the trend we
restrict ourselves to the linear dependence on [O] only.
Oxygen abundances of the galaxies from the LCG sample vary in the
range from 7.52 to 8.47 with the median value of 8.06.
We note that the use of nonlinear dependences on [O], for example,
$10^{\rm [O]}$, does not make any additional progress in reducing
of the regression RMS standard deviations.

First, we discuss potentially misleading methods of studying the trends in the
dependence of $L$ on [O]. The simplest approach is to calculate the mean $L$
for the subsamples with low and high oxygen abundances and to treat
the obtained difference as an dependence on the oxygen abundance.
Adopting the median value [O] = 8.06 as a value dividing galaxies with
low and high oxygen abundances, we find that the subsamples with high [O]
have the mean $L$ (or SFR) values $\sim$ 1.5 times larger than for the
subsamples with low [O].
However, the mean masses of young stellar population in higher-metallicity
subsamples are larger by a factor of $\sim$ 1.5 in comparison with
subsamples with lower metallicity.
Therefore, the differences in SFRs are mostly due to the differences
in masses of the young stellar population and are not directly related to the
differences in the metallicity.

The best way to investigate the direct impact of galaxy metallicity on its
luminosity for the galaxy with the same values of $m$ and $T$ is
to include the metallicity directly in the regression relation. Therefore, we
fit the SFR values for H$\alpha$ emission line using the set of regressors
\begin{equation}\label{eqn:7}
SFR= C_5mf(T)+C_4m^2+C_6m(\rmn{[O]}-\left<\rmn{[O]}\right>)
\end{equation}
with $T_0=3.2$ Myr and $p=0.75$ Myr$^{-1}$.
Here $\left<\rmn{[O]}\right>$ is the mean oxygen abundance of the
sample galaxies. We subtract this value from the galaxy's oxygen abundance
[O] to make the last term in Eq. \ref{eqn:7} practically
orthogonal to the first one and in this way to keep the results of
the Fisher test for the first two regressors.
Using the LSM we calculate the coefficients, their errors and statistical
significances of the regression defined by Eq. \ref{eqn:7}.
For the H$\alpha$ ``regular'' subsample we obtain $\sigma=3.42$,
$C_5=(2.13 \pm 0.05)\times 10^{-7} (F=1654), C_4=(-3.6 \pm
4.2)\times 10^{-18} (F=0.7), C_6=(-4.7 \pm 1.0)\times 10^{-8}
(F=20)$. For the H$\alpha$ ``irregular'' subsample the derived values are
$\sigma=3.8$,
$C_5=(1.94 \pm 0.03)\times10^{-7} (F=4415), C_4=(3.4 \pm
1.6)\times10^{-18} (F=4.9), C_6=(-3.2 \pm 0.7)\times 10^{-8}
(F=23)$. We note the drop of statistical significance of the nonlinear term
with $C_4$ below the threshold for both subsamples.

Switching to the UV luminosities, we generalise Eq. \ref{eqn:4} by adding the
term with the dependence on metallicity:
\begin{equation}\label{eqn:8}
\mathrm{SFR}=C_2m+C_3mT^2+C_4m^2+C_6m(\rmn{[O]}-\left<\rmn{[O]}\right>).
\end{equation}
Derived coefficients, their
errors and statistical significances for FUV and NUV subsamples
are shown in Table \ref{tbl:4}. All coefficients $C_6$ are
negative and have the statistical significanse more than 99.5\%. Coefficients
$C_4$ are statistically insignificant for all FUV and NUV subsamples, similar
to that for H$\alpha$ subsamples.

Summarising, we find that a nonlinear term $m^2$ in the regression
relations (Eqs. \ref{eqn:7} and \ref{eqn:8}) is statistically
insignificant for all six subsamples. Speaking more precisely, we
conclude that the statistical analysis gives us no reason to
justify the existence of such a term.
Would it be statistically significant it makes the ratio $L/m$ be depending on
the young stellar population masses.
This could be treated as an impact of the environment or as a result of
some kind of an interaction of several regions of star formation.
However, it is difficult to analyse these effects statistically because
of the correlation between $m$ and $T$, which would lead to
ambiguous conclusions.

\begin{table*}[tb]
\caption{Values of the coefficients in Eq. \ref{eqn:8}}
\begin{tabular}{lccccccc}
\hline Subsample&$N$&$\sigma$&$\bigstrut C_2\times
10^8$($F$)&$C_3\times 10^{22}$($F$)&$C_4\times 10^{18}$($F$)&$C_6\times
10^{8}$($F$)&$\left<[{\rm O}]\right>$\\ \hline
\multicolumn{8}{c}{a) Regressions with $C_4$ $\ne$ 0} \\ \hline
3.FUV,regular          &$213$&$4.7$&13.6$\pm$0.9(231)&--31.9$\pm$4.1( 60)&$ 30.9\pm10.5(8.6)$&$ -7.0\pm 1.7(16.2)$&8.05\\
4.FUV,irregular        &$418$&$4.4$&10.2$\pm$0.5(375)&--20.7$\pm$2.7( 60)&$ -3.4\pm3.5(1.0)$&$ -3.9\pm 1.0(16.7)$&8.13\\
5.NUV,regular          &$233$&$5.5$&16.0$\pm$1.0(246)&--30.2$\pm$4.7( 42)&$ 23.8\pm12.1(3.9)$&$ -5.8\pm 1.9( 9.1)$&8.05\\
6.NUV,irregular        &$435$&$7.3$&17.8$\pm$0.9(405)&--39.1$\pm$4.3( 83)&$ 12.9\pm5.6(5.3)$&$-12.0\pm 1.6(57.7)$&8.13\\ \hline
\multicolumn{8}{c}{b) Regressions with $C_4$ $=$ 0} \\\hline
3.FUV,regular          &$213$&$4.7$&13.8$\pm$0.9(231)&--26.7$\pm$3.8( 50)&0&$ -4.9\pm 1.6( 9.2)$&8.05\\
4.FUV,irregular        &$418$&$4.4$&10.3$\pm$0.5(378)&--22.3$\pm$2.2(101)&0&$ -3.8\pm 0.9(16.1)$&8.13\\
5.NUV,regular          &$233$&$5.5$&16.1$\pm$1.0(248)&--26.2$\pm$4.2( 38)&0&$ -4.2\pm1.7( 5.7)$&8.05\\
6.NUV,irregular        &$435$&$7.3$&17.3$\pm$0.9(403)&--33.5$\pm$3.6( 88)&0&$-12.2\pm 1.6(59.9)$&8.13\\
7.FUV,all              &$631$&$4.8$&11.7$\pm$0.5(575)&--25.2$\pm$2.0(157)&0&$ -4.8\pm 0.9(31.7)$&8.10\\
8.NUV,all              &$668$&$6.8$&17.1$\pm$0.7(647)&--31.6$\pm$2.8(130)&0&$ -9.7\pm1.2(65.4)$&8.10\\ \hline
\end{tabular}\label{tbl:4}
\end{table*}

\begin{figure}[tb]
\includegraphics[width=\columnwidth]{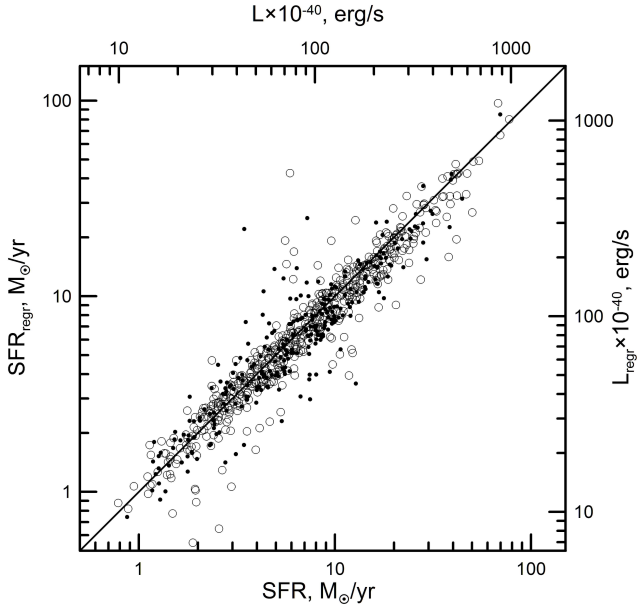}
\caption{Luminosities in the $\rmn{H}\alpha$ line obtained from the regression
relation Eq. \ref{eqn:9} with $p= 0.75$ Myr$^{-1}$. vs. the measured ones. Dots and open circles
correspond to subsamples of galaxies with ``regular'' and ``irregular'' shape}
\label{fig:4}
\end{figure}

We discard nonlinear term with $m^2$ adopting $C_4$ = 0 in Eq. \ref{eqn:7}
and obtain a regression relation in the form
\begin{equation}\label{eqn:9}
SFR = C_5 m f(T)+C_6 m (\rmn{[O]}-\left<\rmn{[O]}\right>).
\end{equation}
We apply this relation for the H$\alpha$ radiation.
For the subsamples No.1 and 2 we use the function $f(T)$ from Eq. \ref{eqn:5}
with $T_0=3.2$ Myr and $p=0.75$ Myr$^{-1}$.
The values, errors and the
statistical significances of the coefficients obtained by the LSM are shown
in Table \ref{tbl:5} (case (a)). Using these coefficients
we calculate SFR$_{regr}$(H$\alpha$) and $L_{regr}$(H$\alpha$) =
SFR$_{regr}$(H$\alpha$)/$k$ for every galaxy
from these subsamples, where $k$ = $7.9 \times 10^{-42}$.
The comparison of the calculated values with the
observed ones is plotted in Figure \ref{fig:4}.
It follows from the Figure that Eq.
\ref{eqn:9} provides a good approximation of the observational
H$\alpha$ data in the entire range of SFR(H$\alpha$) = 0.8 -- 77
$M_{\odot}$ yr$^{-1}$.

\begin{table*}[tb]
\caption{Values of the coefficients in Eq. \ref{eqn:9}}
\begin{tabular}{lcccccc}
\hline Subsample&$N$&$\sigma$&$p$, Myr$^{-1}$&$\bigstrut C_5\times 10^{7}$($F$)
&$C_6\times 10^{8}$($F$)&$\left<[{\rm O}]\right>$\\ \hline
\multicolumn{7}{c}{a) Regressions with $p=0.75$ Myr$^{-1}$} \\ \hline
1.H$\alpha$, regular    &$276$&$3.4$&0.75&2.10$\pm$0.04(2752)&--4.8$\pm$1.0(21)&8.05\\
2.H$\alpha$, irregular  &$519$&$3.8$&0.75&1.98$\pm$0.02(7156)&--3.4$\pm$0.7(27)&8.13\\  \hline
\multicolumn{7}{c}{b) Regressions with the optimal $p$} \\\hline
1.H$\alpha$, regular    &$276$&$3.3$&0.66&1.95$\pm$0.04(2913)&--4.6$\pm$1.0(21)&8.05\\
2.H$\alpha$, irregular  &$519$&$3.6$&0.65&1.82$\pm$0.02(7849)&--3.2$\pm$0.6(26)&8.13\\
9.H$\alpha$, all        &$795$&$3.5$&0.65&1.85$\pm$0.02(10851)&--3.6$\pm$0.5(46)&8.10\\
7.FUV, all              &$631$&$4.7$&0.42&1.11$\pm$0.02(2771)&--5.1$\pm$0.8(39)&8.10\\
8.NUV, all              &$668$&$6.7$&0.32&1.60$\pm$0.03(3925)&--10.1$\pm$1.1(79)&8.10\\ \hline
\multicolumn{7}{c}{c) Regressions with the optimal $p$ and alternative correction for extinction} \\\hline
7.FUV, all              &$631$&$4.7$&0.43&1.12$\pm$0.02(2799)&--5.2$\pm$0.8(41)&8.10\\
8.NUV, all              &$668$&$6.7$&0.33&1.63$\pm$0.03(3981)&--10.3$\pm$1.1(83)&8.10\\ \hline
\end{tabular}\label{tbl:5}
\end{table*}

For the FUV and NUV subsamples we use Eq. \ref{eqn:8} with $C_4=0$.
We can rewrite it in the form of Eq. \ref{eqn:9}, introducing
\begin{equation}\label{eqn:10}
f(T)=1-\eta  T^2,\ C_5=C_2,\ \eta=-\frac{C_3}{C_2}.
\end{equation}
From the two last rows of Table \ref{tbl:4} (case (b)) we derive
$\eta=(21.6\pm  0.9)\times 10^{-3}$ Myr$^{-2}$ for the FUV band and
$\eta=(18.5\pm  0.9)\times 10^{-3}$ Myr$^{-2}$ for the NUV band. Certainly, the
relation Eq. \ref{eqn:10} cannot be used if the starburst age $T$ is greater
than 7 Myr because $f(T)$ becomes negative. Note that the errors of $\eta$
were estimated taking into account not only the errors of $C_2$ and $C_3$
in Table \ref{tbl:4} but their covariation too by using all elements of the
correlation matrix including the nondiagonal ones.

Consider the final fine tuning of the parameters in the Eq.
\ref{eqn:9}. Using LSM we obtain the optimal values of $T_0$ and
$p$ in Eq. \ref{eqn:5}. For the H$\alpha$ ``regular'' subsample we
derive $\sigma=3.3$, $T_0=3.3$ Myr, $p=0.69$, while for the
H$\alpha$ ``irregular'' subsample the derived values are
$\sigma=3.6$, $T_0=3.3$ Myr, $p=0.68$.

Could Eq. \ref{eqn:9} be used for the UV continuum radiation
with $f(T)$ in the form of Eq. \ref{eqn:5}, similar to the H$\alpha$ radiation?
Such attempt turns out to be successful.
For the subsamples No.7 and 8 we obtain the optimal values $T_0=2.9$ Myr and
$T_0=3.1$ Myr, respectively. They are smaller than 3.2 Myr, probably, due to
the larger data scatter in comparison with the H$\alpha$ subsamples.
With these optimal values of $T_0$ and $p$, the RMS standard deviations
for the UV bands are slightly decreased as compared to the case when $f(T)$ is
used in the form of Eq. \ref{eqn:10}.
The minimum of the sum of square residuals $\sum ({\rm SFR}_i-{\rm SFR})^2$,
corresponding to the best value of $T_0$, is rather shallow.
Therefore, we can adopt a single
value $T_0=3.2$ Myr for all subsamples.

In Table \ref{tbl:5} (case (b)) we show the final values of the
parameters for the subsamples No. 1, 2, 7, 8 and 9 in the form according to
Eqs. \ref{eqn:9} and \ref{eqn:5}.

One can see that the approximation Eq. \ref{eqn:5} is much better than
Eq. \ref{eqn:10} for the FUV band and is slightly better for the NUV band.
Moreover, the use of Eq. \ref{eqn:5} for $f(T)$ in
Eq. \ref{eqn:9} is more preferable not only because of the decrease
of the RMS standard deviation $\sigma$, but also by the
same dependence on $T$ as that in the case of H$\alpha$ radiation.
Starting from $T_0$, the fading half-times of the H$\alpha$, FUV and NUV
emission are 1.1 Myr, 1.6 Myr and 2.1 Myr, respectively.

It is noted in Sect. \ref{s:Ext} that for the reddening correction
of galaxy fluxes we
use $E(B-V)_{\rm SDSS}$ derived from the hydrogen Balmer decrement.
Alternatively, we also consider reddening corrections, adopting
$E(B-V)_{\rm SDSS}$ if $E(B-V)_{\rm SDSS}$ $>$ $E(B-V)_{\rm NED}$
(for $\sim$ 90\% of the sample)
and $E(B-V)_{\rm NED}$ if $E(B-V)_{\rm SDSS}$ $<$ $E(B-V)_{\rm NED}$
($\sim$ 10\% of the sample).
The coefficients in this case are shown in Table \ref{tbl:5}, (case (c)).
The comparison of case (b) and case (c) coefficients shows that
differences are very small, indicating that both approaches
can equally be used.

\section{Star formation rates}\label{s:SFR}

We already noted in Section \ref{s:Dep} that there are
different indicators of star formation in a wide range of wavelengths
from UV to radio and different calibrations to quantify it.
As it is mentioned in numerous papers, each SFR
indicator possesses its own strengths and disadvantages. Recently,
the hybrid SFR indices were proposed, which are based on the combination of
the ultraviolet and infrared tracers, the H$\alpha$ and the infrared or radio
continuum tracers, the [O {\sc ii}] $\lambda$3727\AA\ forbidden-line doublet
and the infrared or radio continuum tracers. Studies of star formation
rates for different samples of galaxies with different level of
star formation activity and with different SF tracers were carried out
in many papers
\citep[see, e.g. ][]{Bo,C10,Gil10,Hop02,IP06,IP08,K98,K2009,K02,L2009,Li,M06,O09,SW,Sch}.

In the present paper, the galaxy luminosities in the H$\alpha$ emission line
and in the UV non-ionising continuum are used to obtain SFRs
(Eq. \ref{eqn:1}). The H$\alpha$ emission in the star-forming galaxies
is produced by the gas ionised by the most massive short-lived hot O-stars
with masses $ \gtrsim$ 17 $M_\odot$ and traces the star formation over
the period of a few Myr, corresponding to the lifetime of these stars.
The non-ionising UV emission is produced by stars in a wider range
of masses and therefore can in principle be used as a SFR tracer on a
time scale of up to 100 Myr. However, in the case of strongly
star-forming LCGs, similar trends in Figs.
\ref{fig:2} and \ref{fig:3} imply that H$\alpha$, FUV and NUV emission in LCGs
are produced by the same young stellar populations. This conclusion is
supported by the fact that instantaneous burst with the age of 6 Myr emits
$\sim$ 4 times and $\sim$ 3 times less radiation in the FUV and NUV ranges,
respectively, as compared to that in the burst with the age of 3 Myr
\citep{L99}. Similar difference is seen in Fig. \ref{fig:3}.
Calculating SFRs from Eq. \ref{eqn:1} we actually
use the galaxy luminosities observed at a certain current moment. However, in
star-bursting galaxies, the observed H$\alpha$ and UV-luminosities depend on
the burst age and may vary on a time scale of several Myr.
This effect is most pronounced for the H$\alpha$ luminosity: it is constant
over first $\sim$ 3 Myr of a starburst and then
quickly declines with time. To take into account the temporal luminosity
evolution we introduce the initial
value of the H$\alpha$ luminosity $L_0({\rm H}\alpha)\equiv L({\rm H}\alpha)(T=0)$ after the onset of star formation and calculate the value of
SFR$_0({\rm H}\alpha)=k\times L_0({\rm H}\alpha)$.

Similarly, we also introduce the initial luminosities $L_0$(FUV) and $L_0$(NUV) in the
FUV and NUV ranges and the corresponding values SFR$_0$(FUV) and SFR$_0$(NUV) according to
Eq. \ref{eqn:1}. Hereafter we consider the temporal evolution in UV ranges in
the form of Eq. \ref{eqn:5}. To distinguish the functions $f(T)$ for the H$\alpha$ line and the FUV and NUV ranges we will
use the corresponding subscripts. These functions differ only in the terms of
the coefficient $p$ values, which are presented in the Table \ref{tbl:5},
case (b).
As an illustrative example, we will demonstrate now the certain advantage of
using SFR$_0({\rm H}\alpha)$ in comparison with SFR(H$\alpha$).

\begin{figure}[tb]
\includegraphics[width=\columnwidth]{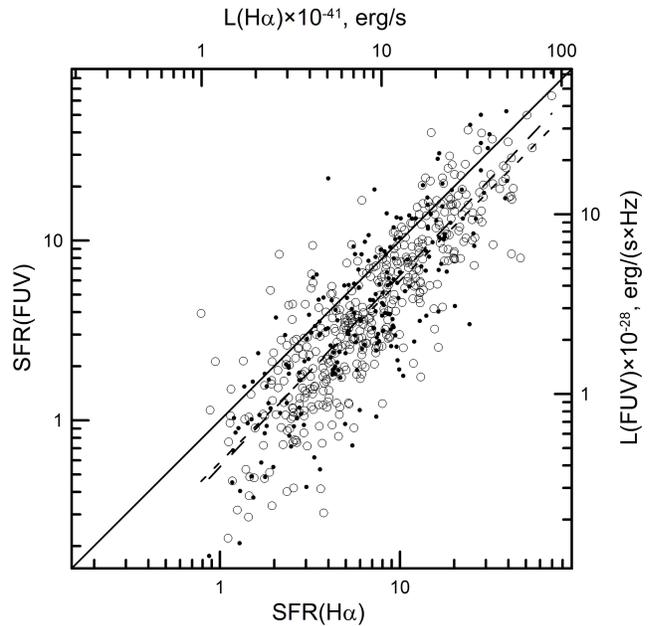}
\caption{Current luminosities and SFRs in the FUV range vs.
luminosities and SFRs in the H$\alpha$ emission line. Dots and
open circles correspond to subsamples of ``regular'' and
``irregular'' galaxies, respectively. Solid line is the line of
equal SFRs, dashed lines show the best fits for subsamples}
\label{fig:5}
\end{figure}

\begin{figure}[tb]
\includegraphics[width=\columnwidth]{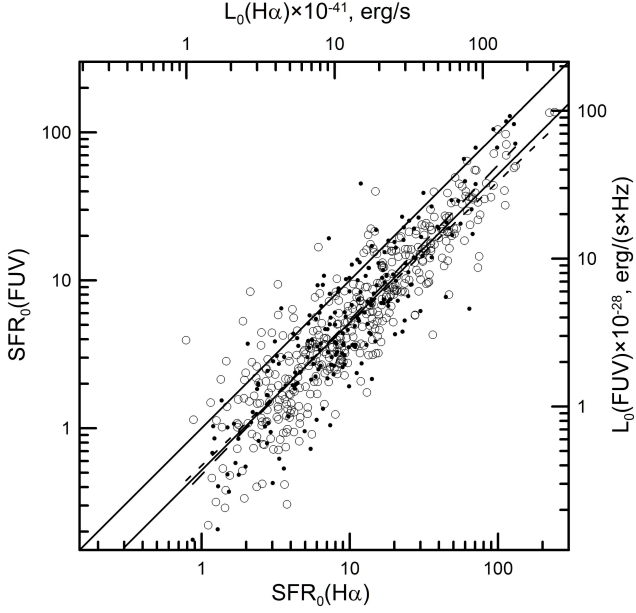}
\caption{Initial luminosities and SFR$_0$ in the FUV range at zero
starburst age vs. initial luminosities and SFR$_0$ in the
H$\alpha$ emission line. Dots and open circles correspond to
subsamples of ``regular'' and ``irregular'' galaxies,
respectively. Upper solid line is the line of equal initial SFRs,
dashed lines show the best fits for subsamples, lower solid line
SFR$_0$(FUV)$=0.52\times$SFR$_0$(H$\alpha$) shows their mean
ratio} \label{fig:6}
\end{figure}

We have an ample sample of the galaxies with known H$\alpha$ and FUV
luminosities. First, we derive SFRs from the observed luminosities.
We show in Figure \ref{fig:5} the relation between SFR(FUV) and
SFR(H$\alpha$) (or equivalently $L$(FUV) and
$L$(H$\alpha$)). It is seen that both SFR(H$\alpha$) and
SFR(FUV) are of the same orders.
The values of the LSM slopes in the dependences
log $L$(FUV) on log $L$(H$\alpha$) for subsamples No. 3 and 4 are
$1.07\pm 0.05$ and $1.04\pm 0.04$ (see dashed lines in Figure
\ref{fig:5}). However, the values of the slopes in the inverse dependences
log $L$(H$\alpha$) on log $L$(FUV) are much smaller than unity,
$0.66\pm 0.03$ and $0.72\pm 0.03$ for subsamples No. 3 and 4, respectively.
These values would correspond to the lines with the slopes 1/0.66 = 1.51 and
1/0.72 = 1.40 in Figure \ref{fig:5}.
The differences in the slopes of the direct and inverse dependences
are likely due to the Malmquist bias caused
by our selection of only galaxies with high $L($H$\beta$)
$\geq$ $3\times 10^{40}$ erg s$^{-1}$.

We reduce $L$s and SFRs to a zero age of a starburst according to
\begin{equation}\label{eqn:11}
L_0=L/f(T),\, \  \mathrm{SFR}_0=\mathrm{SFR}/f(T)
\end{equation}
with $f_{H\alpha}(T)$ for H$\alpha$ emission line and
$f_{\mathrm{FUV}}(T)$ for FUV radiation. Adopting the values of $p$ from
Table \ref{tbl:5}, case b), we obtain the
distribution of galaxies shown in Figure \ref{fig:6}.
The LSM gives the slopes $1.04\pm 0.04$ and $0.95\pm 0.03$
for ``regular'' and ``irregular'' galaxies respectively, implying that the
initial galaxy FUV luminosity $L_0$(FUV)
is proportional to the initial H$\alpha$ luminosity $L_0$(H$\alpha$).
Slopes of the inverse linear dependences log SFR$_0$(H$\alpha$) on
log SFR$_0$(FUV) for the same subsamples are $0.76\pm 0.03$ and $0.85\pm 0.03$.
These values correspond to the lines with the slopes
1/0.76 = 1.31 and 1/0.85 = 1.18 in Figure \ref{fig:6}. Thus, though
Malmquist bias is also present for the data reduced to the zero starburst age,
its effect is much smaller because of smaller differences between the slopes
of the direct log SFR$_0$(FUV) - log SFR$_0$(H$\alpha$) and inverse
log SFR$_0$(H$\alpha$) - log SFR$_0$(FUV) dependences.
We also note that the data point scatter in Figure \ref{fig:6}
is slightly smaller than that in Figure \ref{fig:5}.

However, there is a downward shift of the data points
relative to the line of equal SFRs in Figure \ref{fig:6} indicating that
SFRs obtained from different indicators are proportional, but not equal.

\citet{I11} derived a single value of reddening $E(B-V)$ for both the
gaseous and stellar emission assuming uniform distribution of dust.
However, dust in galaxies is distributed non-uniformly. E.g. \citet{C94}
and \citet{CF00} suggested that young massive stars responsible for
the H$\alpha$ emission are located in more dusty regions as compared to
the stars which produce non-ionising UV radiation, including FUV and NUV
ranges. In particular, \citet{C94} assumed that non-ionising UV radiation
is produced by older stars which were formed in regions different from
those where most massive young stars are present.

Could the non-uniform distribution of dust explain the downward
shift in Fig. \ref{fig:6}? Apparently, not. Assuming that $E(B-V)$ for FUV
and NUV ranges is smaller we obtain lower FUV and NUV luminosities. Therefore,
the downward shift would be larger. Furthermore,
as it was already noted above, similar trends in Figs.
\ref{fig:2} and \ref{fig:3} imply that H$\alpha$, FUV and NUV emission in LCGs
are produced by the same young stellar populations, contrary to assumption
by \citet{C94}.

In order to equalise the values of SFR$_0$(H$\alpha$) and SFR$_0$(FUV) the
coefficients $k$ in Eq. \ref{eqn:1} should be changed from their nominal
values. We find that the mean ratios of SFR$_0$(H$\alpha$) /SFR$_0$(FUV) and
SFR$_0$(H$\alpha$) /SFR$_0$(NUV) are equal to 1.9 and 1.5, respectively.
More precisely, we find that
$10^{\left<\log\left({\rm SFR}_0{\rm (FUV)/SFR}_0({\rm H}\alpha)\right)\right>}$ = 0.53,
$10^{\left<\log\left({\rm SFR}_0{\rm (NUV)/SFR}_0({\rm H}\alpha)\right)\right>}$ = 0.69.

Using these average ratios we obtain statistical relations
\begin{equation}\label{eqn:12}
\begin{array}{l}
L_0({\rm FUV})=3.0\times 10^{-14}\times L_0({\rm H}\alpha),\\
L_0({\rm NUV})=3.9\times 10^{-14}\times L_0({\rm H}\alpha),\\
L_0({\rm NUV})=1.3\times L_0({\rm FUV}).
\end{array}
\end{equation}
We adopt the factor $k^*=1.4\times 10^{-28}\times \gamma$ in Eq.
\ref{eqn:1} for the NUV range and derive
the values of modified factors $k^*=5.3\times 10^{-42}\times \gamma$ for the
H$\alpha$ emission line and $k^*=1.8\times
10^{-28}\times \gamma$ for the FUV range. These values correspond to the rough
equality of the SFR$_0$ obtained from the initial H$\alpha$, FUV and
NUV luminosities.
The multiplier $\gamma$ can be used for overall tuning of the set of factors.
It is equal to $\sim$ 1, if the modified coefficient $k^*$(NUV)
is set to its nominal value $k$(NUV) by \citet{K98}.
On the other hand, if $k^*$(H$\alpha$) is set to its nominal value
$k$(H$\alpha$) by \citet{K98}, then $\gamma$ $\sim$ 1.5. Thus, we obtain
estimations of SFR$_0$
\begin{equation}\label{eqn:13}
\begin{array}{l}
{\rm SFR}_0=5.4\times 10^{-42}\times \gamma \times L({\rm H}\alpha) / f_{H\alpha}(T)\\
\phantom{{\rm SFR}_0}=1.8\times 10^{-28}\times \gamma \times L({\rm FUV}) / f_{FUV}(T)\\
\phantom{{\rm SFR}_0}=1.4\times 10^{-28}\times \gamma \times L({\rm NUV}) / f_{NUV}(T),
\end{array}
\end{equation}
where $L$(H$\alpha)$,
$L$(UV) and SFR are measured in erg s$^{-1}$, erg s$^{-1}$ Hz$^{-1}$ and
$M_{\odot}$ yr$^{-1}$, respectively. The relations in Eq. \ref{eqn:13}
give the approximately equal SFR$_0$(H$\alpha$), SFR$_0$(FUV)
and SFR$_0$(NUV) for LCGs from our sample. If the alternative
correction for extinction is used, we obtain very similar results. Only one
coefficient changes and the first row
in Eq. \ref{eqn:13} gets the form
${\rm SFR}_0=5.3\times 10^{-42}\times \gamma \times L({\rm H}\alpha) / f_{H\alpha}(T)$.

What value of the parameter $\gamma$ is preferable? We can measure the flux from galaxies 
in some wavelength ranges but we cannot directly measure their SFRs. These values can be 
estimated using different indicators of star formation. The values of the coefficients $k$ 
in Eq. \ref{eqn:1} for different wavelength ranges are obtained by modeling and therefore 
dependant from the parameters and assumptions in these models. Thus they can vary in some 
intervals. We use the certain values of $k$ for the H$\alpha$ line and the UV range. The 
values SFR$_0$(H$\alpha$) and SFR$_0$(FUV) are in general proportional, but not equal, as 
they must be. Thus we have to use some agreed set of the coefficients $k$ for different 
ranges which matches various estimations of SFR$_0$. We obtain the agreement condition 
for our sample in Eq. \ref{eqn:12}. There is interval of the parameter $\gamma$ in which 
all coefficients $k$ agree with values obtained from modeling. The increase of the quality 
of modeling will lead to fine-tuning of the value of $\gamma$. Unfortunately if we use the 
values of the coefficients in Eq. \ref{eqn:1} for different wavelength ranges we get some 
discrepancy in SFRs obtained. This means that the models used to obtain these values have 
to be improved. In particular it would be useful to search out the ratio of initial values 
of the luminosities in different wavelength ranges to SFR after the onset of star formation.

\citet{L2009} studied the consistency between the SFRs derived from the
FUV continuum and H$\alpha$ emission for a sample of
the dwarf star-forming galaxies. Particularly, authors discuss the dependence
of the number ratio of ionising to non-ionising
photons in the radiation of dwarf galaxies on its metallicity.
We investigated the dependence of both $L$(H$\alpha$)/$L$(FUV) and
$L_0$(H$\alpha$)/$L_0$(FUV) on [O] and do not find any
statistically significant trend. All values of Fisher coefficients
do not exceed $F=3.5$. However, we note that the rather small
range of [O] for the galaxies from our sample makes it difficult
to study this dependence.

Using Eq. \ref{eqn:1} we derive SFRs and obtain the distributions of
SFR(H$\alpha$), SFR(FUV), SFR(NUV) as well as SRF$_0$(H$\alpha$), SFR$_0$(FUV) and SFR$_0$(NUV).
SFRs derived from the luminosities in the H$\alpha$ emission line, FUV and
NUV continuum
vary in the wide ranges $0.8 \div 77$ $M_{\odot}$ yr$^{-1}$, $0.18 \div 86$
$M_{\odot}$ yr$^{-1}$ and $0.24 \div 113$ $M_{\odot}$ yr$^{-1}$, respectively.
The corresponding median values of SFRs are 6.7 $M_{\odot}$ yr$^{-1}$,
3.8 $M_{\odot}$ yr$^{-1}$ and 5.2 $M_{\odot}$ yr$^{-1}$.
The median values of SFR$_0$(H$\alpha$), SFR$_0$(FUV) and SFR$_0$(NUV)
are 8.7 $M_{\odot}$  yr$^{-1}$,
5.1 $M_{\odot}$  yr$^{-1}$ and 6.5 $M_{\odot}$  yr$^{-1}$, respectively.
For comparison, the median value of SFR(H$\alpha$) is
0.92 $M_{\odot}$ yr$^{-1}$ for a sample of about 7000 star-forming galaxies
from the SDSS DR4 (galaxies being less luminous in H$\beta$ than LCGs) \citep{IP08}. \citet{C09} derived mean SFR $\sim$10  $M_{\odot}$ yr$^{-1}$
for the sample of ``green peas''.

Star formation rates SFRs derived from the H$\alpha$, FUV and NUV luminosities
are in better mutual agreement if Eq. \ref{eqn:13} is used instead of Eq.
\ref{eqn:1}. We derive the median values of
$5.8\times \gamma \times M_{\odot} \rmn{ yr}^{-1}$ for SFR$_0$(H$\alpha$),
$6.6\times \gamma \times M_{\odot} \rmn{ yr}^{-1}$ for SFR$_0$(FUV),
and $6.5\times \gamma \times M_{\odot} \rmn{ yr}^{-1}$ for SFR$_0$(NUV).

We already mentioned in Section \ref{s:Dep} that in general Eq. \ref{eqn:1} for
SFR can be applied for the continuous star formation during certain time
interval $\Delta t$. Formally, for strongly star-bursting galaxies, we may
estimate $\Delta t$ as well, assuming the continuous star formation with the
constant SFR and adopting nominal values for the coefficient $k$. Then, it
is expressed as $\Delta t$ = 1/$C_5$ with $C_5$ from Table \ref{tbl:5} and
attains the values in the range $\sim$ 5.1 -- 8.9 Myr for different samples of
case (b).

\begin{figure}[tb]
\includegraphics[width=\columnwidth]{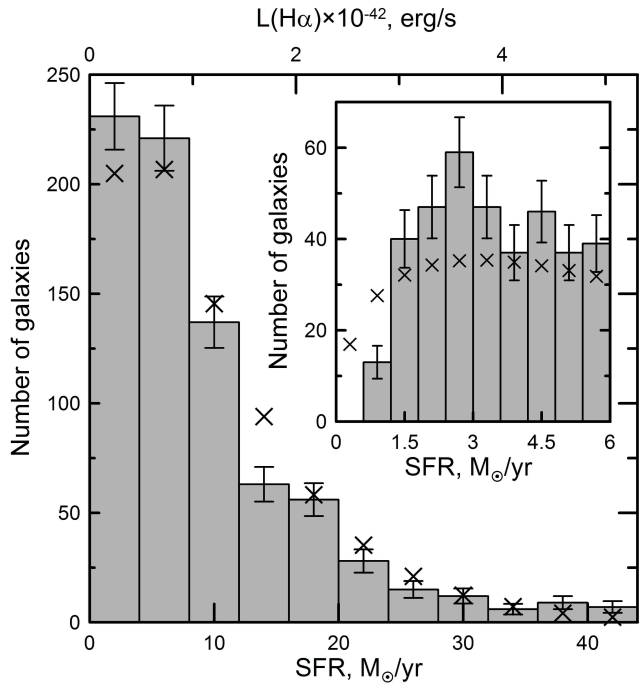}
\caption{Distribution of luminosities $L$(H$\alpha$) and SFR(H$\alpha$).
Diagonal crosses show the distribution
calculated from the Schechter luminosity function with parameters
obtained from the maximum likelihood method. Error bars show the
Poisson errors} \label{fig:7}
\end{figure}

The histogram for the H$\alpha$ emission line luminosity
$L$(H$\alpha$) is shown in Figure
\ref{fig:7}. The standard way to study the luminosity function is its
approximation by the Schechter function \citep{ref:Schechter} in the form
\begin{equation}\label{eqn:14a}
\psi(L)dL=const(L/L^*)^\alpha exp(-L/L^*)d(L/L^*)
\end{equation}
where $\psi$ is the number of galaxies per unit volume in the luminosity interval
from $L$ to $L+dL$. Assuming that the volume $V$ of the galaxies with luminosity
$L$ entering the sample $V\propto L^{3/2}$ and using the maximum likelihood method we obtain
the values $\alpha=-1.04$ and $L^*=8.5\times 10^{41}\,\textrm{erg\,s}^{-1}$.
This value of $L^*$ corresponds to
$\mathrm{SFR}^*=6.7 M_{\odot} \rmn{ yr}^{-1}$ according to Eq. \ref{eqn:1}.
The value $\alpha=-1.04$ is in agreement with the value obtained
from 147,986 galaxy redshifts and fluxes from the SDSS \citep{Bl}.

The distribution of galaxies calculated with
the Schechter function is also shown in Figure \ref{fig:7}.
One can see that there is some difference between real and modelled
distributions of H$\alpha$ luminosities $L$(H$\alpha$).
In particular, model underpredicts the number of
galaxies with largest luminosities calculated from Eq. \ref{eqn:14a}.
The distribution of the observed luminosities in the range of low values
near $10^{41}$ erg s$^{-1}$ is somewhat above the modelled one
(see an inset in Figure \ref{fig:7}). According to the Pearson's $\chi^2$ test
this difference
has the statistical significance over $99\%$. 

The distribution of $L_0$(H$\alpha$) differs from the
Schechter function in a larger extent compared to
the distribution of $L$(H$\alpha$) because of the larger luminosities.
The distributions of the FUV and NUV luminosities are similar to that in
Figure \ref{fig:7} after the luminosity scaling in accordance with the
difference of the median values.

\section{Summary}\label{s:Sum}

We analyse the properties of H$\alpha$ and UV radiation for the sample of
about 800 luminous compact galaxies (LCGs) selected by \citet{I11}
from the Data Release 7 (DR7) of the Sloan Digital Sky Survey (SDSS).
These low-metallicity galaxies are characterised by an active star formation
with the star formation rate SFR(H$\alpha$) in the range
$\sim$ 1 -- 80 $M_\odot$ yr$^{-1}$ and can be considered as local counterparts
of the high-redshift ($z$ $>$ 2) star-forming Lyman-break galaxies (LBGs) and
Ly-$\alpha$ emission-line galaxies (LAEs). We use the optical SDSS
spectroscopic data for LCGs to derive the luminosity in the H$\alpha$ emission
line and SFR(H$\alpha$). {\sl Galaxy Evolution Explorer} ({\sl GALEX})
UV fluxes are used for obtaining luminosities and SFRs in the far-UV
(FUV, $\lambda_{\rm eff}$ = 1528\AA) and in the near-UV
(NUV, $\lambda_{\rm eff}$ = 2271\AA) ranges. These data are supplemented by
other global LCG characteristics derived by \citet{I11} from their
SDSS spectra: chemical element abundances of the interstellar medium, masses
$m$ and ages $T$ of young starbursts. Our main results are as follows:

1. We study the extinction in a sample of LCGs.
It is found that LCGs are rather unobscured galaxies with a mean reddening
about of $E(B-V)=0.136$. The mean internal reddening in the sample
is 0.106. For subsamples of "regular" galaxies with round shape
and "irrregular" ones having shape with some sign of disturbed
morphology we obtain internal reddening of 0.081 and 0.120. The
difference of internal reddening for subsamples is statistically
significant value.

2. We find that the ratio $L$(H$\alpha$)/$m$ in starbursts with
ages $T$ $<$ 3.2 Myr is constant implying that H$\alpha$
luminosity in young starbursts is proportional to the mass of the
young stellar population. At later starburst ages $T$ $\geq$ 3.2
Myr, the ratio $L$(H$\alpha$)/$m$ is declined exponentially with
$T$.  This temporal dependence of the $L$(H$\alpha$)/$m$ ratio is
in general agreement with that from the population synthesis
models by \citet{SV98} which predict the decrease of
$L$(H$\alpha$) after $\sim$ 3 Myr, the lifetime of the most
massive stars.

The dependences of the luminosities per unit mass of the young
stellar population $L$(H$\alpha$)/$m$,
$L$(FUV)/$m$ and $L$(NUV)/$m$ on $T$ (Eqs. \ref{eqn:5},
\ref{eqn:9}) are similar implying that H$\alpha$, FUV and NUV radiation
is produced by the same young populations.  However, the dependences of
$L$(FUV)/$m$ and
$L$(NUV)/$m$ on $T$ are weaker as compared to $L$(H$\alpha$)/$m$.
Starting from $T_0\sim 3.2$
Myr, the half-times of the H$\alpha$, FUV and NUV luminosities
decline are 1.1 Myr, 1.6 Myr and 2.1 Myr, respectively. The ratios
$L$(H$\alpha)/L$(FUV) and $L$(H$\alpha)/L$(NUV) also start to
decrease after $\sim 3.2$ Myr. Thus, these ratios can be used
for estimation of the starburst age $T$.

With this value of $T$ we can estimate $m$ from
$L$(H$\alpha$) in young starbursts without invoking modelling of
spectral energy distribution (SED).
For that we introduce a function $f_{H\alpha}$($T$) which takes
into account the variation of $L$(H$\alpha$)/$m$ with $T$. Then
$m$ $\sim$ $L$(H$\alpha$)/$f_{H\alpha}$($T$).

3. The main impact of galaxy metallicity on its luminosity is the
indirect one through the variation of the mass of the young
stellar population $m$. Dividing the sample of galaxies into
subsamples with high and low metallicities, we obtain that the
mean luminosity will be greater for the subsample with high
metallicity due to the considerable increase of the mean value of
$m$. On the other hand, the direct impact of metallicity has the
opposite sign. The ratio $L/m$ slightly decreases with increasing
of the galaxy metallicity if the starburst age is constant. Thus,
the galaxy luminosity decreases with increasing metallicity at
fixed values of $m$ and the starburst age $T$. This direct impact
is weaker than the indirect one, but it is statistically
significant.

4. Luminosities in H$\alpha$ and UV decrease rapidly
after the starburst age of $\sim 3.2$ Myr.
We take into account this temporal evolution and introduce time-independent
characteristics of the star formation activity, namely the initial
luminosities $L_0$  at the starburst age $T=0$.
The initial luminosities in the H$\alpha$
emission line, FUV and NUV ranges can be obtained from the current
luminosities and the starburst age $T$ from Eq. \ref{eqn:11}.
We find that $L_0$(FUV) and $L_0$(NUV) are proportional to
$L_0$(H$\alpha$) over a large range of luminosities. We can obtain
the approximative equality of the values SFR$_0$ derived from the
initial H$\alpha$, FUV and NUV luminosities by tuning the factor
$k$ in Eq \ref{eqn:1}. The set of factors $k$ for H$\alpha$
emission line and FUV and NUV ranges providing such equality for
the sample of LCGs is used in Eq. \ref{eqn:13}.

5. We find that SFRs derived from the extinction-corrected
H$\alpha$, FUV and NUV luminosities vary in the wide ranges of
0.8$\div$77 $M_\odot$ yr$^{-1}$, 0.18$\div$86 $M_\odot$ yr$^{-1}$
and 0.24$\div$113 $M_\odot$ yr$^{-1}$, respectively. The
corresponding median values of SFRs are 6.7 $M_\odot$ yr$^{-1}$,
3.8 $M_\odot$ yr$^{-1}$ and 5.2 $M_\odot$ yr$^{-1}$. The median
values of initial SFRs are SFR$_0$(H$\alpha$)=8.7 $M_\odot$
yr$^{-1}$, SFR$_0$(FUV)=5.1 $M_\odot$ yr$^{-1}$ and
SFR$_0$(NUV)=6.5 $M_\odot$ yr$^{-1}$. In all cases the nominal coefficients
$k$(H$\alpha$), $k$(FUV) and $k$(NUV) by \citet{K98} are adopted.
The corresponding equalised median SFR$_0$ values in accordance with
Eq. \ref{eqn:13} are equal to
$5.8\times \gamma \times M_{\odot} \rmn{ yr}^{-1}$ for SFR$_0$(H$\alpha$),
$6.6\times \gamma \times M_{\odot} \rmn{ yr}^{-1}$ for SFR$_0$(FUV),
and $6.5\times \gamma \times M_{\odot} \rmn{ yr}^{-1}$ for SFR$_0$(NUV).

\acknowledgments

We thank the anonymous referee for valuable comments which helped
to improve the presentation of results.

This research has made use of the NASA/IPAC Extragalactic Database
(NED) which is operated by the Jet Propulsion Laboratory,
California Institute of Technology, under contract with the
National Aeronautics and Space Administration.

Funding for the Sloan Digital Sky Survey (SDSS) and SDSS-II has
been provided by the Alfred P. Sloan Foundation, the Participating
Institutions, the National Science Foundation, the U.S. Department
of Energy, the National Aeronautics and Space Administration, the
Japanese Monbukagakusho, and the Max Planck Society, and the
Higher Education Funding Council for England.

\end{document}